\documentclass{IEEEtran}

\usepackage[noadjust]{cite}
\usepackage{CJK, amsmath, amssymb}

\newtheorem{lemma}{Lemma}
\newtheorem{theorem}{Theorem}
\newtheorem{conjecture}{Conjecture}
\newtheorem{corollary}{Corollary}

\newtheorem{definition}{Definition}

\DeclareMathOperator{\tr}{tr}
\DeclareMathOperator*{\E}{\mathbb E}

\begin{document}

\title{Entanglement Dynamics From Random Product States: Deviation From Maximal Entanglement}

\begin{CJK}{UTF8}{gbsn}

\author{Yichen Huang (黄溢辰)
\thanks{This work was supported by NSF grant PHY-1818914 and a Samsung Advanced Institute of Technology Global Research Partnership. An earlier version of this paper was presented at the 2021 IEEE International Symposium on Information Theory \cite{Hua21ISIT}.}
\thanks{The author is with the Center for Theoretical Physics, Massachusetts Institute of Technology, Cambridge, MA 02139 USA (e-mail: yichuang@mit.edu).}
}

\maketitle

\end{CJK}

\begin{abstract}

We study the entanglement dynamics of quantum many-body systems and prove the following: (I) For any geometrically local Hamiltonian on a lattice, starting from a random product state the entanglement entropy is bounded away from the maximum entropy at all times with high probability. (II) In a spin-glass model with random all-to-all interactions, starting from any product state the average entanglement entropy is bounded away from the maximum entropy at all times. We also extend these results to any unitary evolution with charge conservation and to the Sachdev-Ye-Kitaev model. Our results highlight the difference between the entanglement generated by (chaotic) Hamiltonian dynamics and that of random states, for the latter is nearly maximal.

\end{abstract}


\section{Introduction}

Entanglement, a concept of quantum information theory, has been widely used in condensed matter and statistical physics to provide insights beyond those obtained via ``conventional'' quantities. A large body of literature is available on the static \cite{HLW94, VLRK03, RM04, Has07, HM14, KLW15, VR17, DLL18, LCB18, LG19, Hua21NPB, AKLV13, Hua14aAl, Hua17aPe} and dynamical \cite{CC05, ZPP08, BPM12, KH13, Hua17AP, GLQ17, Hua21aA} behavior of entanglement in various systems. The scaling of entanglement \cite{ECP10} reflects the classical simulability of quantum many-body systems \cite{Vid03, VC06, SWVC08, Osb12, Hua15T, LVV15, GHLS15, CF16, GE16NJP, DB19, Hua19QV, Hua20ISIT, Hua21JCP, Hua14aAp, Hua15aC, SV17, Hua19aA}.

The dynamics of chaotic (not necessarily geometrically) local Hamiltonians is of high current interest. Since these models are almost by definition not exactly solvable, heuristic descriptions of the universal aspects of the dynamics have been developed. It is important to understand the extent to which the heuristic descriptions reflect reality.

In this paper, we study the entanglement dynamics \cite{LQ19, Hua21PP} starting from a random product state, which is typically a ``massive'' superposition of energy eigenstates \cite{HH19}. The time evolution under a chaotic local Hamiltonian is so complex that heuristically, one might expect that the state at long times behaves like a random state. Therefore,

\begin{conjecture} \label{conj}
For chaotic (not necessarily geometrically) local Hamiltonians, starting from a random product state the entanglement entropy approaches that of a random state at long times.
\end{conjecture}

This conjecture is not mathematically precise because ``chaotic'' is not defined. We do not attempt to define it here, for there is no clear-cut definition of quantum chaos.

Recall that the entanglement entropy of a random state is nearly maximal \cite{Pag93, FK94, San95, Sen96}. However, we prove the following results.
\begin{itemize}
    \item For any geometrically local Hamiltonian on a lattice, starting from a random product state the entanglement entropy is bounded away from the maximum entropy at all times with high probability.
    \item In a spin-glass model with random all-to-all interactions, starting from any product state the average entanglement entropy is bounded away from the maximum entropy at all times.
\end{itemize}
We also extend these results to any unitary evolution with charge conservation and to the Sachdev-Ye-Kitaev (SYK) model \cite{SY93, Kit15, MS16}. Our results highlight the difference between the entanglement generated by (chaotic) Hamiltonian dynamics and that of random states. The difference is a consequence of energy conservation, which prevents the time-evolved state from behaving like a completely random state \cite{HBZ19}. For chaotic Hamiltonian dynamics at long times, if our upper bounds on the entanglement entropy are tight, then the difference is a subleading correction, and Conjecture \ref{conj} holds to leading order.

\section{Preliminaries}

Throughout this paper, standard asymptotic notations are used extensively. Let $f,g:\mathbb R^+\to\mathbb R^+$ be two functions. One writes $f(x)=O(g(x))$ if and only if there exist constants $M,x_0>0$ such that $f(x)\le Mg(x)$ for all $x>x_0$; $f(x)=\Omega(g(x))$ if and only if there exist constants $M,x_0>0$ such that $f(x)\ge Mg(x)$ for all $x>x_0$; $f(x)=\Theta(g(x))$ if and only if there exist constants $M_1,M_2,x_0>0$ such that $M_1g(x)\le f(x)\le M_2g(x)$ for all $x>x_0$.

\begin{definition} [entanglement entropy]
The entanglement entropy of a bipartite pure state $\rho_{AB}$ is defined as the von Neumann entropy
\begin{equation}
S(\rho_A):=-\tr(\rho_A\ln\rho_A)
\end{equation}
of the reduced density matrix $\rho_A=\tr_B\rho_{AB}$.
\end{definition}

We briefly review the entanglement of random states.

\begin{theorem} [conjectured and partially proved by Page \cite{Pag93}; proved in Refs. \cite{FK94, San95, Sen96}] \label{pagethm}
For a bipartite pure state $\rho_{AB}$ chosen uniformly at random with respect to the Haar measure,
\begin{equation} \label{page}
\E_{\rho_{AB}}S(\rho_A)=\sum_{k=d_B+1}^{d_Ad_B}\frac{1}{k}-\frac{d_A-1}{2d_B}=\ln d_A-\frac{d_A}{2d_B}+\frac{O(1)}{d_Ad_B},
\end{equation}
where $d_A\le d_B$ are the local dimensions of subsystems $A$ and $B$, respectively.
\end{theorem}

Let $\gamma\approx0.577216$ be the Euler-Mascheroni constant. The second step of Eq. (\ref{page}) uses the formula
\begin{equation}
\sum_{k=1}^{d_B}\frac{1}{k}=\ln d_B+\gamma+\frac{1}{2d_B}+O(1/d_B^2).
\end{equation}

The distribution of $S(\rho_A)$ is highly concentrated around the mean $\E_{\rho_{AB}}S(\rho_A)$ \cite{HLW06}. This can be seen from the exact formula \cite{VPO16, Wei17} for the variance of $S(\rho_A)$.

Consider a system of $N$ qubits labeled by $1,2,\ldots,N$. Let
\begin{equation}
\sigma_j^x=
\begin{pmatrix}
0 & 1 \\
1 & 0
\end{pmatrix},\quad
\sigma_j^y=
\begin{pmatrix}
0 & -i \\
i & 0
\end{pmatrix},\quad
\sigma_j^z=
\begin{pmatrix}
1 & 0 \\
0 & -1
\end{pmatrix}
\end{equation}
be the Pauli matrices for qubit $j$.

\begin{definition} [Haar-random product state] \label{def:haar}
Let $|\Psi\rangle=\bigotimes_{j=1}^N|\Psi_j\rangle$ be a Haar-random product state, where each $|\Psi_j\rangle$ is chosen independently and uniformly at random with respect to the Haar measure.
\end{definition}

\section{Results}

This section consists of four independent subsections, which can be read without consulting each other.

\subsection{Geometrically local Hamiltonians} \label{ss:r1}

For notational simplicity and without loss of generality, we present the results for geometrically local Hamiltonians in one spatial dimension. (It is easy to see that the same result holds in higher dimensions.) Consider a chain of $N$ qubits governed by a local Hamiltonian
\begin{equation}
H^\textnormal{lat}=\sum_{j=1}^NH_j,
\end{equation}
where $H_j$ represents the nearest-neighbor interaction between qubits at positions $j$ and $j+1$. For concreteness, we use periodic boundary conditions, but our argument also applies to other boundary conditions. Assume without loss of generality that $\tr H_j=0$ (traceless) so that the mean energy of $H^\textnormal{lat}$ is $0$. We do not assume translational invariance. In particular, $\|H_j\|$ may be site dependent but should be $\Theta(1)$ for all $j$.

Let $A$ be a contiguous subsystem of $n$ qubits and $\bar A$ be the rest of the system. Assume without loss of generality that $n\le N/2$. Let $\E_{|A|=n}$ denote averaging over all contiguous subsystems of size $n$. There are $N$ such subsystems.

\begin{theorem} \label{thm:lat}
Initialize the system in a Haar-random product state $|\Psi\rangle$ (Definition \ref{def:haar}). Let
\begin{equation}
\rho_A(t)=\tr_{\bar A}(e^{-iH^\textnormal{lat}t}|\Psi\rangle\langle\Psi|e^{iH^\textnormal{lat}t})
\end{equation}
be the reduced density matrix of subsystem $A$ at time $t$. For $n>1$,
\begin{equation}
\Pr_{\Psi}\left(\sup_{t\in\mathbb R}\E_{|A|=n}S(\rho_A(t))= n\ln2-\Omega(n/N)\right)\ge1-\delta,
\end{equation}
where $\delta>0$ is an arbitrarily small constant.
\end{theorem}

\begin{corollary} \label{cor}
Using the notation of Theorem \ref{thm:lat}, if $H^\textnormal{lat}$ is translationally invariant, then for $n>1$,
\begin{equation} \label{eq:upper}
\sup_{t\in\mathbb R}\E_{\Psi}S(\rho_A(t))=n\ln2-\Omega(n/N).
\end{equation}
\end{corollary}

\begin{IEEEproof}
Since the ensemble of Haar-random product states is translationally invariant, averaging over subsystems is not necessary if we average over this ensemble.
\end{IEEEproof}

For $1<n=O(1)$, the bound (\ref{eq:upper}) is saturated by any translationally invariant $H^{\textnormal{lat}}$ whose spectrum has non-degenerate gaps.

\begin{definition} [non-degenerate gap] \label{def:ndg}
The spectrum $\{E_j\}$ of a Hamiltonian has non-degenerate gaps if the differences $\{E_j-E_k\}_{j\neq k}$ are all distinct, i.e., for any $j\neq k$,
\begin{equation} \label{eq:ndg}
E_j-E_k=E_{j'}-E_{k'}\implies(j=j')~\textnormal{and}~(k=k').
\end{equation}
\end{definition}

\begin{theorem} \label{thm:TI}
Using the notation of Theorem \ref{thm:lat}, if $H^\textnormal{lat}$ is translationally invariant and if the spectrum has non-degenerate gaps, then for $1\le n=O(1)$ and sufficiently large $\tau$,
\begin{equation} \label{eq:TI}
\Pr_{t\in[0,\tau]}\left(\E_{\Psi}S(\rho_A(t))=n\ln2-O(1/N)\right)=1-e^{-\Omega(N)},
\end{equation}
where $t$ is uniformly distributed in the interval $[0,\tau]$.
\end{theorem}

\subsection{Unitary evolution with charge conservation} \label{ss:r2}

Consider a system of $N$ qubits without an underlying lattice structure (of course, Theorem \ref{thm:charge} below remains valid in the presence of a lattice).

Let $m,n$ be positive integers such that $n\le N/2$ and that $mn$ is a multiple of $N$. Let $A_1,A_2,\ldots,A_m$ be $m$ possibly overlapping subsystems, each of which has exactly $n$ qubits. Suppose that each qubit in the system is in exactly $mn/N$ out of these $m$ subsystems. For each $j$, let $\bar A_j$ be the complement of $A_j$ so that $A_j\otimes\bar A_j$ defines a bipartition of the system.

Let $\sigma^z:=\sum_{j=1}^N\sigma_j^z$ be the total charge operator and $U(t)$ be a unitary operator such that $[U(t),\sigma^z]=0$. Note that $U(t)$ need not be generated by a time-independent Hamiltonian. It can be the time evolution operator of a quantum circuit with charge conservation \cite{RPv19, ZL20, Hua20IOPSN, Hua19aD}.

\begin{theorem} \label{thm:charge}
Initialize the system in a Haar-random product state $|\Psi\rangle$ (Definition \ref{def:haar}). Let
\begin{equation}
\rho_{A_j}(t)=\tr_{\bar A_j}(U(t)|\Psi\rangle\langle\Psi|U^\dag(t))
\end{equation}
be the reduced density matrix of subsystem $A_j$ at time $t$. Then,
\begin{equation}
\Pr_{\Psi}\left(\sup_{t\in\mathbb R}\frac1m\sum_{j=1}^mS(\rho_{A_j}(t))= n\ln2-\Omega(n/N)\right)\ge1-\delta,
\end{equation}
where $\delta>0$ is an arbitrarily small constant.
\end{theorem}

\subsection{Spin-glass model} \label{ss:r3}

Consider a system of $N$ qubits. Let $J:=\{J_{jklm}\}_{1\le j<k\le N}^{l,m\in\{x,y,z\}}$ be a collection of $d_N:=9N(N-1)/2$ independent real Gaussian random variables with zero mean $\overline{J_{jklm}}=0$ and unit variance $\overline{J_{jklm}^2}=1$. The Hamiltonian of the spin-glass model is \cite{ES14}
\begin{equation} \label{eq:sg}
H^\textnormal{sg}_J=\frac1{\sqrt{d_N}}\sum_{1\le j<k\le N}\sum_{l,m\in\{x,y,z\}}J_{jklm}\sigma_j^{l}\sigma_k^{m}.
\end{equation}

Let $A\subset\{1,2,\ldots,N\}$ so that $A\sqcup\bar A$ defines a bipartition of the system. Assume without loss of generality that $|A|\le N/2$. Let $\E_{|A|=n}$ denote averaging over all subsystems of size $n$. There are $\binom{N}n$ such subsystems.

\begin{theorem} \label{thm:all}
Initialize the system in an arbitrary (deterministic) product state $|\psi\rangle=\bigotimes_{j=1}^N|\psi_j\rangle$. Let
\begin{equation}
\rho_{J,A}(t_J)=\tr_{\bar A}(e^{-iH^\textnormal{sg}_Jt_J}|\psi\rangle\langle\psi|e^{iH^\textnormal{sg}_Jt_J})
\end{equation}
be the reduced density matrix of subsystem $A$ at time $t_J$. For $n>1$,
\begin{equation} \label{eq:main2}
\E_J\sup_{t_J\in\mathbb R}\E_{|A|=n}S(\rho_{J,A}(t_J))=n\ln2-\Omega(n^2/N^2).
\end{equation}
\end{theorem}

\subsection{SYK model} \label{ss:r4}

Consider a system of $N$ Majorana fermions $\chi_1,\chi_2,\ldots,\chi_N$ with $\{\chi_j,\chi_k\}=2\delta_{jk}$, where $N$ is an even number. Let $K:=\{K_{jklm}\}_{1\le j<k<l<m\le N}$ be a collection of $\binom{N}4$ independent real Gaussian random variables with zero mean $\overline{K_{jklm}}=0$ and unit variance $\overline{K_{jklm}^2}=1$. The Hamiltonian of the SYK model is \cite{SY93, Kit15, MS16}
\begin{equation} \label{eq:syk}
H^\textnormal{SYK}_K=\frac{1}{\sqrt{\binom{N}4}}\sum_{1\le j<k<l<m\le N}K_{jklm}\chi_j\chi_k\chi_l\chi_m.
\end{equation}

Let $A\subset\{1,2,\ldots,N\}$ with $|A|$ even so that $A\sqcup\bar A$ defines a bipartition of the system. Assume without loss of generality that $|A|\le N/2$. Let $\E_{|A|=n}$ denote averaging over all subsystems of size $n$. There are $\binom{N}n$ such subsystems.

\begin{theorem} \label{thm:syk}
Initialize the system in a state $|\psi\rangle$ such that a constant fraction of the expectation values $\{\langle\psi|\chi_j\chi_k\chi_l\chi_m|\psi\rangle\}_{1\le j<k<l<m\le N}$ are non-vanishing, i.e.,
\begin{equation} \label{eq:engf}
|\{(j,k,l,m):|\langle\psi|\chi_j\chi_k\chi_l\chi_m|\psi\rangle|=\Theta(1)\}|=\Theta(N^4).
\end{equation}
Let
\begin{equation}
\rho_{K,A}(t_K)=\tr_{\bar A}(e^{-iH^\textnormal{SYK}_Kt_K}|\psi\rangle\langle\psi|e^{iH^\textnormal{SYK}_Kt_K})
\end{equation}
be the reduced density matrix of subsystem $A$ at time $t_K$. For $n\ge4$,
\begin{equation}
\E_K\sup_{t_K\in\mathbb R}\E_{|A|=n}S(\rho_{K,A}(t_K))=\frac{n\ln2}{2}-\Omega(n^4/N^4).
\end{equation}
\end{theorem}

Unfortunately, not all product states satisfy Eq. (\ref{eq:engf}). It is not difficult to see that the product states defined in Ref. \cite{KM17} are counterexamples. One might expect that a Haar-random product state, if properly defined, satisfies Eq. (\ref{eq:engf}) with overwhelming probability.

In fermionic systems, defining a Haar-random product state is tricky. Since the Hamiltonian (\ref{eq:syk}) conserves fermion parity, the Hilbert space is split into an even sector and an odd sector, which do not interact with each other. It is controversial whether to allow the superposition of states from both sectors. While being compatible with the axioms of quantum mechanics, such a superposition is widely believed to be unphysical. On the other hand, it is not clear how to define a Haar-random product state with definite fermion parity. The statement of Theorem \ref{thm:syk} avoids the controversy and related technical difficulties by introducing the condition (\ref{eq:engf}) instead of claiming $|\psi\rangle$ to be a Haar-random product state.

\section{Proofs}

This section consists of four subsections. Subsections \ref{ss:p1}, \ref{ss:p2}, \ref{ss:p3}, \ref{ss:p4} use the notations of Subsections \ref{ss:r1}, \ref{ss:r2}, \ref{ss:r3}, \ref{ss:r4}, respectively.

\subsection{Proof of Theorem \ref{thm:lat}} \label{ss:p1}

\begin{lemma} \label{l:subadd}
For a (possibly mixed) density matrix $\rho$, let $\rho_A=\tr_{\bar A}\rho$ be the reduced density matrix of subsystem $A$. For $n>1$,
\begin{equation} \label{eq:l1}
    \E_{|A|=n}S(\rho_A)\le\frac{n}2\E_{|A|=2}S(\rho_A).
\end{equation}
\end{lemma}

\begin{IEEEproof}
Using the subadditivity \cite{AL70} of the von Neumann entropy,
\begin{gather}
\E_{|A|=n}S(\rho_A)\le\E_{|A|=n-2}S(\rho_A)+\E_{|A|=2}S(\rho_A),\\
\E_{|A|=1}S(\rho_A)\ge\frac12\E_{|A|=2}S(\rho_A).
\end{gather}
Using the strong subadditivity \cite{LR73},
\begin{equation}
    \E_{|A|=3}S(\rho_A)\le2\E_{|A|=2}S(\rho_A)-\E_{|A|=1}S(\rho_A).
\end{equation}
Combining these inequalities, we obtain (\ref{eq:l1}).
\end{IEEEproof}

\begin{lemma} [\cite{Hua19NPB}] \label{l:2site}
Let $\rho_j$ be a density matrix of qubits at positions $j$ and $j+1$ such that
\begin{equation}
    |\tr(\rho_jH_j)|\ge\epsilon_j\|H_j\|
\end{equation}
for some $\epsilon_j\ge0$. Then,
\begin{equation}
S(\rho_j)\le2\ln2-\epsilon_j^2/2.
\end{equation}
\end{lemma}

\begin{IEEEproof}
We include the proof of this lemma for completeness. Let $I_4$ be the identity matrix of order $4$. Let $\|X\|_1:=\tr\sqrt{X^\dag X}$ denote the trace norm. Since $H_j$ is traceless, $\epsilon_j$ provides a lower bound on the deviation of $\rho_j$ from the maximally mixed state:
\begin{multline}
\epsilon_j\le|\tr(\rho_jH_j)|/\|H_j\|=|\tr((\rho_j-I_4/4)H_j)|/\|H_j\|\\
\le\|\rho_j-I_4/4\|_1=\sum_{i=1}^4|\lambda_i-1/4|,
\end{multline}
where $\lambda_1,\lambda_2,\lambda_3,\lambda_4$ are the eigenvalues of $\rho_j$. An upper bound on $S(\rho_j)$ is $\max\{-\sum_{i=1}^4p_i\ln p_i\}$ subject to the constraints
\begin{equation}
\sum_{i=1}^4p_i=1,\quad\sum_{i=1}^4|p_i-1/4|\ge\epsilon_j.
\end{equation}
Since the Shannon entropy is Schur concave, it suffices to consider the following three cases:
\begin{itemize}
    \item $p_1=p_2=1/4+\epsilon_j/4$, $p_3=p_4=1/4-\epsilon_j/4$;
    \item $p_1=1/4+\epsilon_j/2$, $p_2=p_3=p_4=1/4-\epsilon_j/6$;
    \item (if $\epsilon_j\le1/2$) $p_1=1/4-\epsilon_j/2$, $p_2=p_3=p_4=1/4+\epsilon_j/6$.
\end{itemize}
In all these cases, by Taylor expansion we can prove
\begin{equation}
-\sum_{i=1}^4p_i\ln p_i\le2\ln2-\epsilon_j^2/2
\end{equation}
for $\epsilon_j\ll1$. We have checked numerically that this inequality remains valid for any $\epsilon_j\le1$.
\end{IEEEproof}

\begin{lemma} \label{l:engr}
For a Haar-random product state $|\Psi\rangle$ (Definition \ref{def:haar}),
\begin{equation} \label{eq:engr}
\Pr_{\Psi}(|\langle\Psi|H^\textnormal{lat}|\Psi\rangle|=\Omega(\sqrt N))\ge1-\delta.
\end{equation}
\end{lemma}

\begin{IEEEproof}
For $j=2,3,\ldots,N$, we assume that the expansion of $H_j$ in the Pauli basis does not contain any terms acting only on the qubit at position $j$ (this is without loss of generality since such terms can be included in $H_{j-1}$). Under this assumption, it is easy to see that
\begin{equation}
\E_{\Psi_{j+1}}\langle\Psi_j\Psi_{j+1}|H_j|\Psi_j\Psi_{j+1}\rangle=0
\end{equation}
for any $|\Psi_j\rangle$. Thus, $\{\langle\Psi_j\Psi_{j+1}|H_j|\Psi_j\Psi_{j+1}\rangle\}_{j=1}^{N-1}$ is a martingale difference sequence, and (\ref{eq:engr}) follows from the martingale central limit theorem.
\end{IEEEproof}

We are ready to prove Theorem \ref{thm:lat}. Let
\begin{equation}
    \rho=e^{-iH^\textnormal{lat}t}|\Psi\rangle\langle\Psi|e^{iH^\textnormal{lat}t},\quad\epsilon_j=|\tr(\rho H_j)|/\|H_j\|
\end{equation}
so that
\begin{multline} \label{eq:engc}
    \sum_{j=1}^N\epsilon_j=\sum_{j=1}^N\Theta(|\tr(\rho H_j)|)=\Omega(1)\left|\sum_{j=1}^N\tr(\rho H_j)\right|\\
    =\Omega(|\tr(\rho H^\textnormal{lat})|)=\Omega(|\langle\Psi|H^\textnormal{lat}|\Psi\rangle|).
\end{multline}
Note that $\rho,\epsilon_j$ are functions of time and should carry $t$ as an argument, which is omitted for notational simplicity. Using Lemmas \ref{l:subadd}, \ref{l:2site}, the RMS-AM inequality, and Eq. (\ref{eq:engc}) sequentially,
\begin{multline}
\E_{|A|=n}S(\rho_A)\le\frac{n}2\E_{|A|=2}S(\rho_A)\le\frac{n}{2N}\sum_{j=1}^N(2\ln2-\epsilon_j^2/2)\\
\le n\ln2-\frac{n}{4N^2}\left(\sum_{j=1}^N\epsilon_j\right)^2=n\ln2-\frac{n\Omega(\langle\Psi|H^\textnormal{lat}|\Psi\rangle^2)}{N^2}.
\end{multline}
We complete the proof of Theorem \ref{thm:lat} by combining this inequality with Lemma \ref{l:engr}.

\subsection{Proof of Theorem \ref{thm:charge}} \label{ss:p2}

The following lemmas are analogues of Lemmas \ref{l:subadd}, \ref{l:2site}, \ref{l:engr}, respectively.

\begin{lemma}
For a (possibly mixed) density matrix $\rho$, let $\rho_{A_j}=\tr_{\bar A_j}\rho$ be the reduced density matrix of subsystem $A_j$, and $\rho_k$ be that of qubit $k$. Then,
\begin{equation}
    \frac{1}{m}\sum_{j=1}^mS(\rho_{A_j})\le\frac{n}{N}\sum_{k=1}^NS(\rho_k).
\end{equation}
\end{lemma}

\begin{lemma}
Let $\rho_j$ be a density matrix of qubit $j$. Then,
\begin{equation}
S(\rho_j)\le\ln2-\tr^2(\rho_j\sigma^z_j)/2.
\end{equation}
\end{lemma}

\begin{lemma}
For a Haar-random product state $|\Psi\rangle$,
\begin{equation}
\Pr_{\Psi}(|\langle\Psi|\sigma^z|\Psi\rangle|=\Omega(\sqrt N))\ge1-\delta.
\end{equation}
\end{lemma}

Let $\rho=U(t)|\Psi\rangle\langle\Psi|U^\dag(t)$. Theorem \ref{thm:charge} can be proved in almost the same way as Theorem \ref{thm:lat} by replacing $H_j,H^\textnormal{lat}$ with $\sigma^z_j,\sigma^z$, respectively.

\subsection{Proof of Theorem \ref{thm:all}} \label{ss:p3}

\paragraph*{Proof overview} We observe that all product states satisfy the energy condition (\ref{eq:eng}), which is preserved under time evolution. To obtain an upper bound on the left-hand side of Eq. (\ref{eq:main2}), we maximize the average subsystem entropy subject to the energy constraint (\ref{eq:eng}). Since the thermal state maximizes the von Neumann entropy for a given energy, we assign a temperature to each subsystem for each disorder realization of the Hamiltonian (\ref{eq:sg}). Lemma \ref{l:same} implies that in order to maximize the average subsystem entropy, all these temperatures must have the same absolute value. Finally, we upper bound the average subsystem entropy using the thermodynamic relation (Lemma \ref{l:thermo}) between energy and entropy.

\paragraph*{Complete proof} We start with the spectral and thermodynamic properties of the spin-glass model (\ref{eq:sg}).

\begin{lemma} \label{l:moment}
For any positive integer $k$,
\begin{equation} \label{eq:moment}
\frac1{2^{2N}}\E_J\tr^2((H^\textnormal{sg}_J)^k)\le\frac1{2^N}\E_J\tr((H^\textnormal{sg}_J)^{2k})\le(2k-1)!!.
\end{equation}
\end{lemma}

\begin{IEEEproof}
The first step follows from the RMS-AM inequality. The second step can be proved in the same way as (35) of Ref. \cite{FTW19}.
\end{IEEEproof}

Let
\begin{equation}
    \varrho_J(\beta):=e^{-\beta H^\textnormal{sg}_J}/\tr e^{-\beta H^\textnormal{sg}_J}
\end{equation}
be the thermal state of $H^\textnormal{sg}_J$ at inverse temperature $\beta$. Define a measure on $\mathbb R^{d_N}$ such that
\begin{equation}
    \int_{\mathcal J}\,\mathrm dJ=\Pr(J\in\mathcal J),\quad\forall \mathcal J\subseteq\mathbb R^{d_N}.
\end{equation}
For an arbitrary bipartition of $\mathbb R^{d_N}=\mathcal J^+\sqcup\mathcal J^-$, let
\begin{equation}
    \mathcal E(\beta):=\int_{\mathcal J^+}\tr(\varrho_J(\beta)H^\textnormal{sg}_J)\,\mathrm dJ-\int_{\mathcal J^-}\tr(\varrho_J(-\beta)H^\textnormal{sg}_J)\,\mathrm dJ
\end{equation}
so that $\mathcal E(0)=0$ and that $\mathcal E$ is strictly monotonically decreasing.

\begin{lemma} \label{l:energy}
For $-c\le\beta\le 0$ with a small constant $c=\Theta(1)$,
\begin{equation}
    \mathcal E(\beta)\le-\beta+O(\beta^2).
\end{equation}
\end{lemma}

\begin{IEEEproof}
Since $H^\textnormal{sg}_J$ is traceless,
\begin{equation} \label{eq:par}
    \tr e^{-\beta H^\textnormal{sg}_J}\ge2^N,\quad\forall J.
\end{equation}
Using (\ref{eq:par}), Lemma \ref{l:moment}, and the RMS-AM inequality,
\begin{align}
    &\int_{\mathcal J^+}\tr(\varrho_J(\beta)H^\textnormal{sg}_J)\,\mathrm dJ-\int_{\mathcal J^-}\tr(\varrho_J(-\beta)H^\textnormal{sg}_J)\,\mathrm dJ\nonumber\\
    \le&\int_{\mathcal J^+}\frac{\tr(e^{-\beta H^\textnormal{sg}_J}H^\textnormal{sg}_J)}{2^N}\,\mathrm dJ-\int_{\mathcal J^-}\frac{\tr(e^{\beta H^\textnormal{sg}_J}H^\textnormal{sg}_J)}{2^N}\,\mathrm dJ\nonumber\\
    =&\sum_{k=0}^{+\infty}\int_{\mathcal J^+}\frac{(-\beta)^k\tr((H^\textnormal{sg}_J)^{k+1})}{k!2^N}\,\mathrm dJ\nonumber\\
    &-\sum_{k=0}^{+\infty}\int_{\mathcal J^-}\frac{\beta^k\tr((H^\textnormal{sg}_J)^{k+1})}{k!2^N}\,\mathrm dJ\nonumber\\
    =&-\beta\sum_{k=0}^{+\infty}\frac{\beta^{2k}\E_J\tr((H^\textnormal{sg}_J)^{2k+2})}{(2k+1)!2^N}\nonumber\\
    &+\sum_{k=1}^{+\infty}\frac{\beta^{2k}}{(2k)!2^N}\left(\int_{\mathcal J^+}-\int_{\mathcal J^-}\right)\tr((H^\textnormal{sg}_J)^{2k+1})\,\mathrm dJ\nonumber\\
    \le&-\beta e^{\beta^2/2}+\sum_{k=1}^{+\infty}\frac{\beta^{2k}}{(2k)!2^N}\sqrt{\E_J\tr^2((H^\textnormal{sg}_J)^{2k+1})}\nonumber\\
    \le&-\beta e^{\beta^2/2}+\sum_{k=1}^{+\infty}\frac{\beta^{2k}\sqrt{(4k+1)!!}}{(2k)!}=-\beta+O(\beta^2).
\end{align}
\end{IEEEproof}

Let
\begin{equation}
    \mathcal S(\beta):=\int_{\mathcal J^+}S(\varrho_J(\beta))\,\mathrm dJ+\int_{\mathcal J^-}S(\varrho_J(-\beta))\,\mathrm dJ
\end{equation}
so that $\mathcal S(0)=N\ln2$ and that $\mathcal S$ is strictly monotonically increasing (decreasing) for negative (positive) $\beta$.

\begin{lemma} \label{l:thermo}
For $\beta$ such that $0\le\mathcal E(\beta)=O(1)$,
\begin{equation} \label{eq:ent}
    \mathcal S(\beta)=N\ln2-\Omega((\mathcal E(\beta))^2).
\end{equation}
\end{lemma}

\begin{IEEEproof}
Lemma \ref{l:energy} implies that
\begin{equation}
\beta=-\Omega(\mathcal E(\beta)).
\end{equation}
Combining this with the thermodynamic relation
\begin{equation}
    \mathrm d\mathcal S(\beta)/\mathrm d\beta=\beta\mathrm d\mathcal E(\beta)/\mathrm d\beta\implies\mathrm d\mathcal S(\beta)/\mathrm d\mathcal E(\beta)=\beta,
\end{equation}
we obtain Eq. (\ref{eq:ent}).
\end{IEEEproof}

We are ready to prove Theorem \ref{thm:all}. Recall that $|\psi\rangle=\bigotimes_{j=1}^N|\psi_j\rangle$ is an arbitrary (deterministic) product state. Let
\begin{equation}
\mathcal J_+:=\{J:\langle\psi|H^\textnormal{sg}_J|\psi\rangle>0\},\quad\mathcal J_-:=\{J:\langle\psi|H^\textnormal{sg}_J|\psi\rangle<0\}.
\end{equation}
$\mathcal J_+$ and $\mathcal J_-$ have the same volume as $J\in\mathcal J_+$ if and only if $-J\in\mathcal J_-$. Moreover, the complement of $\mathcal J_+\sqcup\mathcal J_-$ has measure zero. Hence,
\begin{equation} \label{eq:split}
\E_J=\frac{1}{2}\E_{J\in\mathcal J_+}+\frac{1}{2}\E_{J\in\mathcal J_-}.
\end{equation}

\begin{lemma} \label{l:eng}
\begin{equation} \label{eq:eng}
\E_J|\langle\psi|H^\textnormal{sg}_J|\psi\rangle|=\Theta(1).
\end{equation}
\end{lemma}

\begin{IEEEproof}
It follows from observation that 
\begin{multline}
\langle\psi|H^\textnormal{sg}_J|\psi\rangle=1/\sqrt{d_N}\\
\times\sum_{1\le j<k\le N}\sum_{l,m\in\{x,y,z\}}J_{jklm}\langle\psi_j|\sigma_j^l|\psi_j\rangle\langle\psi_k|\sigma_k^m|\psi_k\rangle
\end{multline}
is the sum of $\Theta(N^2)$ independent Gaussian random variables divided by $\Theta(N)$.
\end{IEEEproof}

Let
\begin{equation}
H^\textnormal{sg}_{J,A}=\frac1{\sqrt{d_{|A|}}}\sum_{j,k\in A;j<k}\sum_{l,m\in\{x,y,z\}}J_{jklm}\sigma_j^{l}\sigma_k^{m}.
\end{equation}
Since $\sqrt{d_{|A|}/d_N}H^\textnormal{sg}_{J,A}$ is the restriction of $H^\textnormal{sg}_J$ to subsystem $A$,
\begin{equation} \label{eq:sub}
H^\textnormal{sg}_J=\sqrt{d_N/d_n}\E_{|A|=n}H^\textnormal{sg}_{J,A}\otimes I_{\bar A},
\end{equation}
where $I_{\bar A}$ is the identity operator on $\bar A$. Combining Eq. (\ref{eq:sub}) with Eq. (\ref{eq:split}) and Lemma \ref{l:eng},
\begin{multline} \label{eq:change}
    \E_{J\in\mathcal J_+}\E_{|A|=n}\tr(\rho_{J,A}(t_J)H^\textnormal{sg}_{J,A})\\
    -\E_{J\in\mathcal J_-}\E_{|A|=n}\tr(\rho_{J,A}(t_J)H^\textnormal{sg}_{J,A})=\Theta(n/N).
\end{multline}

An upper bound on the left-hand side of Eq. (\ref{eq:main2}) can be obtained as follows. For each tuple $(J,A)$, we introduce a density matrix $\varrho_{J,A}$ supported on $A$. Since $\rho_{J,A}(t_J)$ and $\varrho_{J,A}$ are not related to each other, we use different fonts for rho to avoid confusion. We maximize $\E_J\E_{|A|=n}S(\varrho_{J,A})$ subject to the constraint
\begin{multline} \label{eq:ec}
        \E_{J\in\mathcal J_+}\E_{|A|=n}\tr(\varrho_{J,A}H^\textnormal{sg}_{J,A})-\E_{J\in\mathcal J_-}\E_{|A|=n}\tr(\varrho_{J,A}H^\textnormal{sg}_{J,A})\\
        =\Theta(n/N).
\end{multline}
Lemma \ref{l:same} below implies that the maximum is achieved when
\begin{equation} \label{eq:thermal}
    \varrho_{J,A}=e^{\mp\beta H^\textnormal{sg}_{J,A}}/\tr e^{\mp\beta H^\textnormal{sg}_{J,A}}
\end{equation}
is the thermal state of $H^\textnormal{sg}_{J,A}$ at inverse temperature $\pm\beta$ for $J\in\mathcal J_\pm$, respectively.

\begin{lemma} \label{l:same}
Let $M$ be a positive integer and $E$ be a real number. For $i=1,2,\ldots,M$, let $G_i$ be a Hamiltonian on the Hilbert space $\mathcal H_i$, and $\varrho_i$ be a density matrix on $\mathcal H_i$. The maximum average entropy $\sum_{i=1}^MS(\varrho_i)/M$ subject to the constraint
\begin{equation} \label{eq:st}
\frac{1}{M}\sum_{i=1}^M\tr(\varrho_iG_i)=E
\end{equation}
is achieved when every $\varrho_i=e^{-\beta G_i}/\tr e^{-\beta G_i}$ is a thermal state at the same temperature, and the inverse temperature $\beta$ can be obtained by solving the constraint (\ref{eq:st}).
\end{lemma}

\begin{IEEEproof}
Let $\varrho:=\bigotimes_{i=1}^M\varrho_i$ be a density matrix on the Hilbert space $\mathcal H:=\bigotimes_{i=1}^M\mathcal H_i$, and
\begin{equation}
G:=\sum_{i=1}^MI^{\otimes (i-1)}\otimes G_i\otimes I^{\otimes (M-i)}
\end{equation}
be a Hamiltonian on $\mathcal H$ so that
\begin{equation}
\tr(\varrho G)=\sum_{i=1}^M\tr(\varrho_iG_i)=ME.
\end{equation}
The von Neumann entropy is additive: $S(\varrho)=\sum_{i=1}^MS(\varrho_i)$. To maximize $S(\varrho)$, $\varrho$ must be a thermal state of $G$ \cite{Weh78}:
\begin{equation}
\varrho=e^{-\beta G}/\tr e^{-\beta G}=\bigotimes_{i=1}^Me^{-\beta G_i}/\tr e^{-\beta G_i}.
\end{equation}
Thus, each $\varrho_i$ is a thermal state of $G_i$ at the same inverse temperature $\beta$.
\end{IEEEproof}

Since $H^\textnormal{sg}_{J,A}$ is traceless, $\tr(e^{-\beta H^\textnormal{sg}_{J,A}}H^\textnormal{sg}_{J,A})$ is positive (negative) for negative (positive) $\beta$. Substituting Eq. (\ref{eq:thermal}) into Eq. (\ref{eq:ec}), we see that the solution $\beta$ is negative. Since $H^\textnormal{sg}_{J,A}$ is a spin-glass Hamiltonian for a system of $n$ spins, Lemma \ref{l:thermo} implies that
\begin{equation} \label{eq:final}
    \E_J\E_{|A|=n}S(\varrho_{J,A})=n\ln2-\Omega((n/N)^2).
\end{equation}
We complete the proof of Theorem \ref{thm:all} by noting that the left-hand side of Eq. (\ref{eq:final}) is an upper bound on $\E_J\E_{|A|=n}S(\rho_{J,A}(t_J))$ for any $\{t_J\in\mathbb R\}_J$.

\subsection{Proof of Theorem \ref{thm:syk}} \label{ss:p4}

Theorem \ref{thm:syk} can be proved in almost the same way as Theorem \ref{thm:all}. As an analogue of Lemma \ref{l:eng},
\begin{equation}
\E_K|\langle\psi|H^\textnormal{SYK}_K|\psi\rangle|=\Theta(1)
\end{equation}
follows from Eq. (\ref{eq:engf}). Moreover, ``$n/N$'' in Eqs. (\ref{eq:change}), (\ref{eq:ec}), (\ref{eq:final}) and ``$n\ln2$'' in Eq. (\ref{eq:final}) should be modified to $n^2/N^2$ and $n(\ln2)/2$, respectively.

\appendix[Proof of Theorem \ref{thm:TI}]

Let $\{|j\rangle\}_{j=1}^{2^N}$ be a complete set of eigenstates of $H^{\textnormal{lat}}$ and $\varrho_{j,A}:=\tr_{\bar A}|j\rangle\langle j|$ be the reduced density matrix of subsystem $A$. The energy basis $\{|j\rangle\}$ is unambiguously defined. This is because the non-degenerate gap condition (\ref{eq:ndg}) implies that all eigenvalues of $H^{\textnormal{lat}}$ are distinct. Recall that $n$ is the number of qubits in $A$.

\begin{lemma} \label{l:lower}
For $n=O(1)$,
\begin{equation}
    \frac1{2^N}\sum_{j=1}^{2^N}S(\varrho_{j,A})=n\ln2-O(1/N).
\end{equation}
\end{lemma}

\begin{IEEEproof}
Using the monotonicity of the R\'enyi entropy and Theorem 1 in Ref. \cite{KLW15},
\begin{align}
    &\frac1{2^N}\sum_{j=1}^{2^N}S(\varrho_{j,A})\ge-\frac{1}{2^N}\sum_{j=1}^{2^N}\ln\tr(\varrho_{j,A}^2)\nonumber\\
    &\ge-\ln\left(\frac{1}{2^N}\sum_{j=1}^{2^N}\tr(\varrho_{j,A}^2)\right)\ge-\ln\left(\frac1{2^n}+\frac{2^n}N\right)\nonumber\\
    &=n\ln2-O(1/N).
\end{align}
\end{IEEEproof}

The effective dimension of $|\Psi\rangle$ is defined as
\begin{equation}
1/D^{\textnormal{eff}}_\Psi=\sum_{j=1}^{2^N}|\langle j|\Psi\rangle|^4.
\end{equation}

\begin{lemma} [\cite{HH19}] \label{l:eff}
\begin{equation}
    \Pr_{\Psi}(D^{\textnormal{eff}}_\Psi=e^{\Omega(N)})=1-e^{-\Omega(N)}.
\end{equation}
\end{lemma}

Let
\begin{equation}
\rho^\infty:=\lim_{\tau\to+\infty}\frac1\tau\int_0^\tau\rho(t)\,\mathrm dt,\quad\rho(t):=e^{-iH^{\textnormal{lat}}t}|\Psi\rangle\langle\Psi|e^{iH^{\textnormal{lat}}t}
\end{equation}
be the infinite time average and $\rho^\infty_A:=\tr_{\bar A}\rho^\infty$ be the reduced density matrix of subsystem $A$. Expanding $|\Psi\rangle$ in the energy basis, it is easy to see that
\begin{equation}
    \rho^\infty=\sum_{j=1}^{2^N}p_j|j\rangle\langle j|,\quad p_j:=|\langle j|\Psi\rangle|^2
\end{equation}
is the so-called diagonal ensemble. Since the spectrum of $H^{\textnormal{lat}}$ has non-degenerate gaps,

\begin{lemma} [\cite{LPSW09, Sho11}] \label{l:eq}
\begin{equation} \label{eq:eq}
\lim_{\tau\to+\infty}\frac1\tau\int_0^\tau\|\rho_A(t)-\rho^\infty_A\|_1\,\mathrm dt\le 2^n\big/\sqrt{D^{\rm eff}_\Psi}.
\end{equation}
\end{lemma}

\begin{lemma} [continuity of the von Neumann entropy \cite{Fan73, Aud07}] \label{l:cont}
Let $T:=\|\rho-\rho'\|_1/2$ be the trace distance between two density matrices $\rho,\rho'$ on the Hilbert space $\mathbb C^D$. Then,
\begin{equation}
    |S(\rho)-S(\rho')|\le T\ln(D-1)-T\ln T-(1-T)\ln(1-T).
\end{equation}
\end{lemma}

Since by definition $0\le T\le1$, the right-hand side of this inequality is well defined.

We are ready to prove Theorem \ref{thm:TI}. Lemmas \ref{l:eff}, \ref{l:eq} imply that
\begin{equation}
\lim_{\tau\to+\infty}\frac1\tau\int_0^\tau\E_{\Psi}\|\rho_A(t)-\rho^\infty_A\|_1\,\mathrm dt=e^{-\Omega(N)}.
\end{equation}
Markov's inequality implies that
\begin{equation} \label{eq:prob}
    \Pr_{t\in[0,\tau]}\left(\E_{\Psi}\|\rho_A(t)-\rho^\infty_A\|_1=e^{-\Omega(N)}\right)=1-e^{-\Omega(N)}
\end{equation}
for sufficiently large $\tau$. Due to the continuity of the von Neumann entropy (Lemma \ref{l:cont}),
\begin{align} \label{eq:close}
    &\E_{\Psi}\|\rho_A(t)-\rho^\infty_A\|_1=e^{-\Omega(N)}\nonumber\\
    \implies&\left|\E_{\Psi}S(\rho_A(t))-\E_{\Psi}S(\rho^\infty_A)\right|\le\E_{\Psi}|S(\rho_A(t))-S(\rho^\infty_A)|\nonumber\\
    &=e^{-\Omega(N)}.
\end{align}
Using the concavity of the von Neumann entropy and Lemma \ref{l:lower},
\begin{multline} \label{eq:last}
    \E_{\Psi}S(\rho^\infty_A)=\E_{\Psi}S\left(\sum_{j=1}^{2^N}p_j\varrho_{j,A}\right)\ge\sum_{j=1}^{2^N}\E_{\Psi}p_jS(\varrho_{j,A})\\
    =\frac1{2^N}\sum_{j=1}^{2^N}S(\varrho_{j,A})=n\ln2-O(1/N).
\end{multline}
Equation (\ref{eq:TI}) follows from (\ref{eq:prob}), (\ref{eq:close}), and (\ref{eq:last}).

\section*{Acknowledgment}

The author would like to thank Yingfei Gu, Daniel H. Ranard, and Shreya Vardhan for asking questions that motivated Theorems \ref{thm:charge}, \ref{thm:TI}, and Corollary \ref{cor}, respectively. The author also thanks Y.G. for collaboration on a related project \cite{HG19} and S.V. for pointing out a typo in a draft of this paper. 

\bibliographystyle{IEEEtran}
\bibliography{main}

\begin{thebibliography}{10}
\providecommand{\url}[1]{#1}
\csname url@samestyle\endcsname
\providecommand{\newblock}{\relax}
\providecommand{\bibinfo}[2]{#2}
\providecommand{\BIBentrySTDinterwordspacing}{\spaceskip=0pt\relax}
\providecommand{\BIBentryALTinterwordstretchfactor}{4}
\providecommand{\BIBentryALTinterwordspacing}{\spaceskip=\fontdimen2\font plus
\BIBentryALTinterwordstretchfactor\fontdimen3\font minus
  \fontdimen4\font\relax}
\providecommand{\BIBforeignlanguage}[2]{{%
\expandafter\ifx\csname l@#1\endcsname\relax
\typeout{** WARNING: IEEEtran.bst: No hyphenation pattern has been}%
\typeout{** loaded for the language `#1'. Using the pattern for}%
\typeout{** the default language instead.}%
\else
\language=\csname l@#1\endcsname
\fi
#2}}
\providecommand{\BIBdecl}{\relax}
\BIBdecl

\bibitem{Hua21ISIT}
Y.~Huang, ``Entanglement dynamics from random product states at long times,''
  in \emph{2021 IEEE International Symposium on Information Theory}, 2021, pp.
  1332--1337.

\bibitem{HLW94}
C.~Holzhey, F.~Larsen, and F.~Wilczek, ``Geometric and renormalized entropy in
  conformal field theory,'' \emph{Nuclear Physics B}, vol. 424, no.~3, pp.
  443--467, 1994.

\bibitem{VLRK03}
G.~Vidal, J.~I. Latorre, E.~Rico, and A.~Kitaev, ``Entanglement in quantum
  critical phenomena,'' \emph{Physical Review Letters}, vol.~90, no.~22, p.
  227902, 2003.

\bibitem{RM04}
G.~Refael and J.~E. Moore, ``Entanglement entropy of random quantum critical
  points in one dimension,'' \emph{Physical Review Letters}, vol.~93, no.~26,
  p. 260602, 2004.

\bibitem{Has07}
M.~B. Hastings, ``An area law for one-dimensional quantum systems,''
  \emph{Journal of Statistical Mechanics: Theory and Experiment}, vol. 2007,
  no.~08, p. P08024, 2007.

\bibitem{HM14}
Y.~Huang and J.~E. Moore, ``Excited-state entanglement and thermal mutual
  information in random spin chains,'' \emph{Physical Review B}, vol.~90,
  no.~22, p. 220202, 2014.

\bibitem{KLW15}
J.~P. Keating, N.~Linden, and H.~J. Wells, ``Spectra and eigenstates of spin
  chain {H}amiltonians,'' \emph{Communications in Mathematical Physics}, vol.
  338, no.~1, pp. 81--102, 2015.

\bibitem{VR17}
L.~Vidmar and M.~Rigol, ``Entanglement entropy of eigenstates of quantum
  chaotic {H}amiltonians,'' \emph{Physical Review Letters}, vol. 119, no.~22,
  p. 220603, 2017.

\bibitem{DLL18}
A.~Dymarsky, N.~Lashkari, and H.~Liu, ``Subsystem eigenstate thermalization
  hypothesis,'' \emph{Physical Review E}, vol.~97, no.~1, p. 012140, 2018.

\bibitem{LCB18}
C.~Liu, X.~Chen, and L.~Balents, ``Quantum entanglement of the
  {S}achdev-{Y}e-{K}itaev models,'' \emph{Physical Review B}, vol.~97, no.~24,
  p. 245126, 2018.

\bibitem{LG19}
T.-C. Lu and T.~Grover, ``Renyi entropy of chaotic eigenstates,''
  \emph{Physical Review E}, vol.~99, no.~3, p. 032111, 2019.

\bibitem{Hua21NPB}
Y.~Huang, ``Universal entanglement of mid-spectrum eigenstates of chaotic local
  {H}amiltonians,'' \emph{Nuclear Physics B}, vol. 966, p. 115373, 2021.

\bibitem{AKLV13}
I.~Arad, A.~Kitaev, Z.~Landau, and U.~Vazirani, ``An area law and
  sub-exponential algorithm for 1{D} systems,'' arXiv:1301.1162.

\bibitem{Hua14aAl}
Y.~Huang, ``Area law in one dimension: Degenerate ground states and {R}\'enyi
  entanglement entropy,'' arXiv:1403.0327.

\bibitem{Hua17aPe}
------, ``Provably efficient neural network representation for image
  classification,'' arXiv:1711.04606.

\bibitem{CC05}
P.~Calabrese and J.~Cardy, ``Evolution of entanglement entropy in
  one-dimensional systems,'' \emph{Journal of Statistical Mechanics: Theory and
  Experiment}, vol. 2005, no.~04, p. P04010, 2005.

\bibitem{ZPP08}
M.~\v{Z}nidari\v{c}, T.~Prosen, and P.~Prelov\v{s}ek, ``Many-body localization
  in the {H}eisenberg {$XXZ$} magnet in a random field,'' \emph{Physical Review
  B}, vol.~77, no.~6, p. 064426, 2008.

\bibitem{BPM12}
J.~H. Bardarson, F.~Pollmann, and J.~E. Moore, ``Unbounded growth of
  entanglement in models of many-body localization,'' \emph{Physical Review
  Letters}, vol. 109, no.~1, p. 017202, 2012.

\bibitem{KH13}
H.~Kim and D.~A. Huse, ``Ballistic spreading of entanglement in a diffusive
  nonintegrable system,'' \emph{Physical Review Letters}, vol. 111, no.~12, p.
  127205, 2013.

\bibitem{Hua17AP}
Y.~Huang, ``Entanglement dynamics in critical random quantum {I}sing chain with
  perturbations,'' \emph{Annals of Physics}, vol. 380, pp. 224--227, 2017.

\bibitem{GLQ17}
Y.~Gu, A.~Lucas, and X.-L. Qi, ``Spread of entanglement in a
  {S}achdev-{Y}e-{K}itaev chain,'' \emph{Journal of High Energy Physics}, vol.
  2017, no.~9, p. 120, 2017.

\bibitem{Hua21aA}
Y.~Huang, ``Adding boundary terms to {A}nderson localized {H}amiltonians leads
  to unbounded growth of entanglement,'' arXiv:2109.07640.

\bibitem{ECP10}
J.~Eisert, M.~Cramer, and M.~B. Plenio, ``Colloquium: Area laws for the
  entanglement entropy,'' \emph{Reviews of Modern Physics}, vol.~82, no.~1, pp.
  277--306, 2010.

\bibitem{Vid03}
G.~Vidal, ``Efficient classical simulation of slightly entangled quantum
  computations,'' \emph{Physical Review Letters}, vol.~91, no.~14, p. 147902,
  2003.

\bibitem{VC06}
F.~Verstraete and J.~I. Cirac, ``Matrix product states represent ground states
  faithfully,'' \emph{Physical Review B}, vol.~73, no.~9, p. 094423, 2006.

\bibitem{SWVC08}
N.~Schuch, M.~M. Wolf, F.~Verstraete, and J.~I. Cirac, ``Entropy scaling and
  simulability by matrix product states,'' \emph{Physical Review Letters}, vol.
  100, no.~3, p. 030504, 2008.

\bibitem{Osb12}
T.~J. Osborne, ``Hamiltonian complexity,'' \emph{Reports on Progress in
  Physics}, vol.~75, no.~2, p. 022001, 2012.

\bibitem{Hua15T}
Y.~Huang, ``Classical simulation of quantum many-body systems,'' Ph.D.
  dissertation, University of California, Berkeley, 2015.

\bibitem{LVV15}
Z.~Landau, U.~Vazirani, and T.~Vidick, ``A polynomial time algorithm for the
  ground state of one-dimensional gapped local {H}amiltonians,'' \emph{Nature
  Physics}, vol.~11, no.~7, pp. 566--569, 2015.

\bibitem{GHLS15}
S.~Gharibian, Y.~Huang, Z.~Landau, and S.~W. Shin, ``Quantum {H}amiltonian
  complexity,'' \emph{Foundations and Trends in Theoretical Computer Science},
  vol.~10, no.~3, pp. 159--282, 2015.

\bibitem{CF16}
C.~T. Chubb and S.~T. Flammia, ``Computing the degenerate ground space of
  gapped spin chains in polynomial time,'' \emph{Chicago Journal of Theoretical
  Computer Science}, vol. 2016, p.~9, 2016.

\bibitem{GE16NJP}
Y.~Ge and J.~Eisert, ``Area laws and efficient descriptions of quantum
  many-body states,'' \emph{New Journal of Physics}, vol.~18, no.~8, p. 083026,
  2016.

\bibitem{DB19}
A.~M. Dalzell and F.~G. S.~L. Brand\~ao, ``Locally accurate {MPS}
  approximations for ground states of one-dimensional gapped local
  {H}amiltonians,'' \emph{{Quantum}}, vol.~3, p. 187, 2019.

\bibitem{Hua19QV}
Y.~Huang, ``Matrix product state approximations: {B}ringing theory closer to
  practice,'' \emph{{Quantum Views}}, vol.~3, p.~26, 2019.

\bibitem{Hua20ISIT}
------, ``2{D} {L}ocal {H}amiltonian with area laws is {QMA}-complete,'' in
  \emph{2020 IEEE International Symposium on Information Theory}, 2020, pp.
  1927--1932.

\bibitem{Hua21JCP}
------, ``Two-dimensional local {H}amiltonian problem with area laws is
  {QMA}-complete,'' \emph{Journal of Computational Physics}, vol. 443, p.
  110534, 2021.

\bibitem{Hua14aAp}
------, ``A polynomial-time algorithm for the ground state of one-dimensional
  gapped {H}amiltonians,'' arXiv:1406.6355.

\bibitem{Hua15aC}
------, ``Computing energy density in one dimension,'' arXiv:1505.00772.

\bibitem{SV17}
N.~Schuch and F.~Verstraete, ``Matrix product state approximations for infinite
  systems,'' arXiv:1711.06559.

\bibitem{Hua19aA}
Y.~Huang, ``Approximating local properties by tensor network states with
  constant bond dimension,'' arXiv:1903.10048.

\bibitem{LQ19}
Y.~D. Lensky and X.-L. Qi, ``Chaos and high temperature pure state
  thermalization,'' \emph{Journal of High Energy Physics}, vol. 2019, no.~6,
  p.~25, 2019.

\bibitem{Hua21PP}
Y.~Huang, ``Extensive entropy from unitary evolution,'' \emph{Preprints}, vol.
  2021, p. 2021040254.

\bibitem{HH19}
Y.~Huang and A.~W. Harrow, ``Scrambling and thermalization in
  translation-invariant systems,'' arXiv:1907.13392.

\bibitem{Pag93}
D.~N. Page, ``Average entropy of a subsystem,'' \emph{Physical Review Letters},
  vol.~71, no.~9, pp. 1291--1294, 1993.

\bibitem{FK94}
S.~K. Foong and S.~Kanno, ``Proof of {P}age's conjecture on the average entropy
  of a subsystem,'' \emph{Physical Review Letters}, vol.~72, no.~8, pp.
  1148--1151, 1994.

\bibitem{San95}
J.~S\'anchez-Ruiz, ``Simple proof of {P}age's conjecture on the average entropy
  of a subsystem,'' \emph{Physical Review E}, vol.~52, no.~5, pp. 5653--5655,
  1995.

\bibitem{Sen96}
S.~Sen, ``Average entropy of a quantum subsystem,'' \emph{Physical Review
  Letters}, vol.~77, no.~1, pp. 1--3, 1996.

\bibitem{SY93}
S.~Sachdev and J.~Ye, ``Gapless spin-fluid ground state in a random quantum
  {H}eisenberg magnet,'' \emph{Physical Review Letters}, vol.~70, no.~21, pp.
  3339--3342, 1993.

\bibitem{Kit15}
A.~Kitaev, ``A simple model of quantum holography,'' in \emph{KITP Program:
  Entanglement in Strongly-Correlated Quantum Matter}, 2015,
  \url{https://online.kitp.ucsb.edu/online/entangled15/kitaev/},
  \url{https://online.kitp.ucsb.edu/online/entangled15/kitaev2/}.

\bibitem{MS16}
J.~Maldacena and D.~Stanford, ``Remarks on the {S}achdev-{Y}e-{K}itaev model,''
  \emph{Physical Review D}, vol.~94, no.~10, p. 106002, 2016.

\bibitem{HBZ19}
Y.~Huang, F.~G. S.~L. Brand\~ao, and Y.-L. Zhang, ``Finite-size scaling of
  out-of-time-ordered correlators at late times,'' \emph{Physical Review
  Letters}, vol. 123, no.~1, p. 010601, 2019.

\bibitem{HLW06}
P.~Hayden, D.~W. Leung, and A.~Winter, ``Aspects of generic entanglement,''
  \emph{Communications in Mathematical Physics}, vol. 265, no.~1, pp. 95--117,
  2006.

\bibitem{VPO16}
P.~Vivo, M.~P. Pato, and G.~Oshanin, ``Random pure states: Quantifying
  bipartite entanglement beyond the linear statistics,'' \emph{Physical Review
  E}, vol.~93, no.~5, p. 052106, 2016.

\bibitem{Wei17}
L.~Wei, ``Proof of {V}ivo-{P}ato-{O}shanin's conjecture on the fluctuation of
  von {N}eumann entropy,'' \emph{Physical Review E}, vol.~96, no.~2, p. 022106,
  2017.

\bibitem{RPv19}
T.~Rakovszky, F.~Pollmann, and C.~W. von Keyserlingk, ``Sub-ballistic growth of
  {R}\'enyi entropies due to diffusion,'' \emph{Physical Review Letters}, vol.
  122, no.~25, p. 250602, 2019.

\bibitem{ZL20}
T.~Zhou and A.~W.~W. Ludwig, ``Diffusive scaling of {R}\'enyi entanglement
  entropy,'' \emph{Physical Review Research}, vol.~2, no.~3, p. 033020, 2020.

\bibitem{Hua20IOPSN}
Y.~Huang, ``Dynamics of {R}\'enyi entanglement entropy in diffusive qudit
  systems,'' \emph{{IOP} {SciNotes}}, vol.~1, no.~3, p. 035205, 2020.

\bibitem{Hua19aD}
------, ``Dynamics of {R}\'enyi entanglement entropy in local quantum circuits
  with charge conservation,'' arXiv:1902.00977.

\bibitem{ES14}
L.~Erd\H{o}s and D.~Schr\"{o}der, ``Phase transition in the density of states
  of quantum spin glasses,'' \emph{Mathematical Physics, Analysis and
  Geometry}, vol.~17, no. 3-4, pp. 441--464, 2014.

\bibitem{KM17}
I.~Kourkoulou and J.~Maldacena, ``Pure states in the {SYK} model and
  nearly-{$AdS_2$} gravity,'' arXiv:1707.02325.

\bibitem{AL70}
H.~Araki and E.~H. Lieb, ``Entropy inequalities,'' \emph{Communications in
  Mathematical Physics}, vol.~18, no.~2, pp. 160--170, 1970.

\bibitem{LR73}
E.~H. Lieb and M.~B. Ruskai, ``Proof of the strong subadditivity of
  quantum-mechanical entropy,'' \emph{Journal of Mathematical Physics},
  vol.~14, no.~12, pp. 1938--1941, 1973.

\bibitem{Hua19NPB}
Y.~Huang, ``Universal eigenstate entanglement of chaotic local
  {H}amiltonians,'' \emph{Nuclear Physics B}, vol. 938, pp. 594--604, 2019.

\bibitem{FTW19}
R.~Feng, G.~Tian, and D.~Wei, ``Spectrum of {SYK} model,'' \emph{Peking
  Mathematical Journal}, vol.~2, no.~1, pp. 41--70, 2019.

\bibitem{Weh78}
A.~Wehrl, ``General properties of entropy,'' \emph{Reviews of Modern Physics},
  vol.~50, no.~2, pp. 221--260, 1978.

\bibitem{LPSW09}
N.~Linden, S.~Popescu, A.~J. Short, and A.~Winter, ``Quantum mechanical
  evolution towards thermal equilibrium,'' \emph{Physical Review E}, vol.~79,
  no.~6, p. 061103, 2009.

\bibitem{Sho11}
A.~J. Short, ``Equilibration of quantum systems and subsystems,'' \emph{New
  Journal of Physics}, vol.~13, no.~5, p. 053009, 2011.

\bibitem{Fan73}
M.~Fannes, ``A continuity property of the entropy density for spin lattice
  systems,'' \emph{Communications in Mathematical Physics}, vol.~31, no.~4, pp.
  291--294, 1973.

\bibitem{Aud07}
K.~M.~R. Audenaert, ``A sharp continuity estimate for the von {N}eumann
  entropy,'' \emph{Journal of Physics A: Mathematical and Theoretical},
  vol.~40, no.~28, pp. 8127--8136, 2007.

\bibitem{HG19}
Y.~Huang and Y.~Gu, ``Eigenstate entanglement in the {S}achdev-{Y}e-{K}itaev
  model,'' \emph{Physical Review D}, vol. 100, no.~4, p. 041901, 2019.

\end{thebibliography}

\end{document}